\documentclass[fp,twocolumn,txfonts]{jpsj3}
\usepackage{graphicx}
\usepackage{color}
\usepackage{dcolumn}
\usepackage{tabularx}
\usepackage{float}
\usepackage{amsmath}

\title{Anomalous infrared spectra of hybridized phonons in type-I clathrate Ba$_8$Ga$_{16}$Ge$_{30}$}
\author{\name{Kei \surname{Iwamoto}}$^1$, \name{Shunsuke \surname{Kushibiki}}$^1$, \name{Hironori} \surname{Honda}$^1$, \name{Shuhei \surname{Kajitani}}$^1$, 
\name{Tatsuya \surname{Mori}}$^1$, \name{Hideki \surname{Matsumoto}}$^1$, \name{Naoki \surname{Toyota}}$^1$, \\
\name{Koichiro \surname{Suekuni}}$^2$, \name{Marcos A. \surname{Avila}}$^{2,4}$, and \name{Toshiro \surname{Takabatake}}$^{2,3}$}
\inst{$^1$Department of Physics, Graduate School of Science, Tohoku University, Sendai 980-8578, Japan\\
$^2$Department of Quantum Matter, ADSM, $^3$Institute for Advanced Material Research, Hiroshima University, Higashi-Hiroshima 739-8530, Japan\\
$^4$Centro de Ci\^encias Naturais e Humanas, Universidade Federal do ABC, Santo Andr\'e 09210-971, Brazil}

\abst{
The optical conductivity spectra of the rattling phonons in the clathrate Ba$_8$Ga$_{16}$Ge$_{30}$ are investigated in detail by use of the terahertz time-domain spectroscopy.
The experiment has revealed that the lowest-lying vibrational mode of a Ba(2)$^{2+}$ ion consists of a sharp Lorentzian peak at 1.2 THz 
superimposed on a broad tail weighted in the lower frequency regime around 1.0 THz. 
With decreasing temperature, an unexpected linewidth broadening of the phonon peak is observed, 
together with monotonic softening of the phonon peak and the enhancement of the tail structure. 
These observed anomalies are discussed in terms of impurity scattering effects on the hybridized phonon system of rattling and acoustic phonons.
}

\kword{Rattling phonons, Hybridized phonon system, Phonon-impurity interactions, THz time-domain spectroscopy, Clathrates}

\begin{document}

\maketitle

\section{Introduction}
A decade or more ago, certain compounds of the clathrates family \cite{Nolas1998,Cohn1999,Sales2001} were found to exhibit metallic electrical conductivity
but a heavily suppressed thermal conductivity resembling glassy materials, which lead to their potential application for thermoelectric devices. 
In general, they have a network structure of atomic cages including guest atoms. 
These guest atoms inside the cages give rise to THz-frequency optical modes, 
which have been observed from the specific heat \cite{Sales2001,Avila2006b,Suekuni2008}, inelastic neutron scattering (INS) \cite{Hermann2005,Christensen2008},
and optical spectroscopy \cite{Nolas2000,Takasu2006,Mori2011,Mori2009} measurements. 
Such a low-frequency mode is considered not only to interfere but also to hybridize strongly with the heat-carrying acoustic phonons of the cage network, 
and thus is expected to play key roles in various anomalous phenomena mentioned above. 
On the basis of such perspective, extensive studies have been conducted on hybridized phonon systems and their related properties.

The type-I clathrate Ba$_8$Ga$_{16}$Ge$_{30}$, hereafter abbreviated as BGG,
has a cubic structure with space group $Pm\bar{3}n$, presenting 2 dodecahedral and 6 tetrakaidecahedral cages formed by a network of Ga and Ge. 
It is a so-called Zintl compound that, for the charge compensation, each cage encapsulates a guest Ba$^{2+}$ ion, labeled Ba(1) and Ba(2), respectively. 
In general, a fine tuning of the Ga/Ge concentration slightly off the stoichiometry turns the system into a heavily doped semiconductor, 
and hence controls both the charge carriers sign and density \cite{Sales2001, Paschen2001, Bentien2004, Avila2006a, Avila2008}.

Each Ba$^{2+}$ ion is loosely bounded to the local potential of an electronegative cage, the tetrakaidecahedron in particular, and therefore vibrates with larger amplitude. 
Such a localized vibrational mode of Ba(2) in the oversized cage is called as {\it rattling phonon} and shows strong anharmonicity. 
This anharmonicity results in a so-called softening whereby the vibrational frequency decreases toward low temperature, 
as demonstrated experimentally by Raman scattering \cite{Takasu2006} and theoretically by a quasiharmonic approximation to a quartic term of the guest ions displacement \cite{Dahm2007}. 

In addition to this anharmonicity, the hybridization with the acoustic phonon is a significant characteristic of the Ba(2)$^{2+}$ rattling vibrations. 
In 2008, INS measurements on BGG have clarified the phonon dispersions at room temperature \cite{Christensen2008}.
A clear evidence is obtained for an {\it avoided crossing} between the rattling vibration of Ba(2) and the cage acoustic modes. 
In comparison to the zone-boundary acoustic phonons of $\sim$ 2\,THz, the rattling phonon with frequency of $\sim$ 1.2\,THz is indeed low enough 
to become hybridized in an avoiding manner such that both phonon modes are strongly coupled and no longer independent of each other. 
It was pointed out that such a hybridization can suppress thermal conductivities at low temperatures \cite{Christensen2008,Nakayama2011}

Under thus background of research, we have been highly motivated to study the charge dynamics of these rattling vibrations, 
and so far we have reported the optical conductivity spectra of BGG \cite{Mori2009} and the isostructural BGS (Ba$_8$Ga$_{16}$Sn$_{30}$) \cite{Mori2011}. 
For the vibration of Ba(2)$^{2+}$ ions, there are two infrared-active normal modes of $T_{1u}$ symmetry, 
labeled as $\nu_{1}$ and $\nu_{2}$ perpendicular and parallel to the local fourfold inversion axis, respectively \cite{Mori2009,Takasu2006}. 
For BGG, the spectral sharpening of the $\nu_1$ mode associated with softening toward low temperature was reported in the previous work \cite{Mori2009}. 
This result was ascribed to both unequally spaced vibrational levels in a single-well anharmonic potential, or the {\it on-center} potential, and the Boltzmann factor, 
expected from the local anharmonic potential model \cite{Matsumoto2009}. 
For BGS, with decreasing temperature, the single broad peak of $\nu_1$ mode became split into two subpeaks and also showed a linewidth broadening. 
The former result was consistently understood by assuming a multi-well anharmonic potential, or the {\it off-center} potential \cite{Matsumoto2009}, 
revealed by structural analyses \cite{Avila2008,Suekuni2008}. 
However, the latter finding was quite unexpected because the low-temperature spectra becomes sharp due to reduction of thermal effects in decay process in general 
and suppression of thermal excitation in the local anharmonic potential model \cite{Matsumoto2009}. 
So far this anomaly has had yet to be explained, 
although we have suggested the relevance of interactions with other excitations such as acoustic phonons and/or charge carriers \cite{Mori2011}.

As mentioned above, the temperature dependence of the linewidth of the rattling phonon spectra seems to show opposite behavior between BGG and BGS; 
the former shows sharpening, while the latter shows broadening with decreasing temperature. 
Then there arises a question whether those contrasting behaviors can be ascribed to the difference of the charge carrier and the potential shapes, 
{\it i. e.}, {\it on-center} or {\it off-center}. 
It is noted that, from the spectra of $p$-type BGG, we can clarify the dynamical properties of the rattling phonon without taking into account any effects of charge carriers, 
that is, we have the information on roles of anharmonicity in the linewidth. 
In fact the dc resistivity increases sharply by a few order of magnitude at low temperature \cite{Avila2006a}, 
indicating that the carrier density and hence the interaction with rattling phonons might be negligibly small. 

Looking back to previous BGG spectra (see Fig. 4 in Ref. \cite{Mori2009}), one finds some spectral weight appearing in the low energy region of the $\nu_1$ peak, 
which may suggest a double-peak structure and may lead to a modification of the analysis. 
In order to have a definite answer on this question, the frequency resolution must be improved much more than in our previous works, 
where the linewidth and the experimental resolution were in the same order. 
Thus we study in more detail the optical conductivity of BGG by improving the resolution in frequency 
and by analyzing carefully a spurious interference effect of multiply reflected light within a sample. 

In the present paper we report newly obtained optical conductivity spectra in the THz range, with an improved frequency resolution over the previous work \cite{Mori2009}. 
The data show that the lowest-lying spectra of the rattling phonons consist of a sharp Lorentzian peak at 1.2\,THz superimposed clearly on a broad tail around 1.0\,THz. 
This paper consists as follows. 
In \S. 2, we describe the experimental procedure and demonstrate how the resolution is improved by using a thicker substrate. 
In \S. 3, we show the optical conductivity spectra where the interference effect is quantitatively taken into account. 
From the temperature dependence of the peak frequencies and linewidths, we conclude that the linewidth broadening appears toward low temperature. 
In \S. 4, we discuss these results by use of the strongly hybridized model of the rattling and acoustic phonons with impurity-type scatterings. 
After commenting that the anharmonicity works as the mean field to induce a frequency softening, 
we suggest that the strong low-frequency hybridization between rattling and acoustic phonons, softening of the rattling phonon, 
and impurity-type scattering of phonons can be causes of the anomalous linewidth broadening toward low temperature.
Finally, in \S. 5 we summarize the present works.

\section{Experimental procedure}
Single crystals with $p$-type carriers were grown by a self-flux method \cite{Avila2006a,Avila2006b}. 
To obtain optimum transmitting signals, disks of about 5\,mm in diameter were {attached with adhesive on a sapphire substrate and} polished down to $d=$10-20\,$\mu$m in thickness. 
Terahertz time-domain spectroscopy (THz-TDS) measurements covering the frequency range of 0.2-3.0\,THz (6.7-100\,cm$^{-1}$) were carried out 
with a Tochigi Nikon RT-20000 spectrometer, which uses a standard technique for the transmission configuration \cite{Mori2008, Mori2009, Mori2011}. 
The optical conductivity is numerically determined by using the equation taking into account multiple reflections in the sample and adhesive. 
Four samples in total were used both for determining the optimum thickness that gives highest signal-to-noise ratio and for more carefully mapping an interference effect. 
By using a thicker sapphire substrate of 2.0\,mm rather than 0.5\,mm in our previous work \cite{Mori2009}, 
the frequency resolution in the present work has been greatly improved from about 100\,GHz to about 25\,GHz. 
For further details on our time-domain spectroscopy, refer also to Refs. \cite{Mori2008, Mori2009, Mori2011}. 
See also Appendix A and B for detailed discussions of an interference effect and a frequency resolution. 

\section{Experimental results}

\begin{figure}[tb]
\begin{center}
\includegraphics[width=7.5cm]{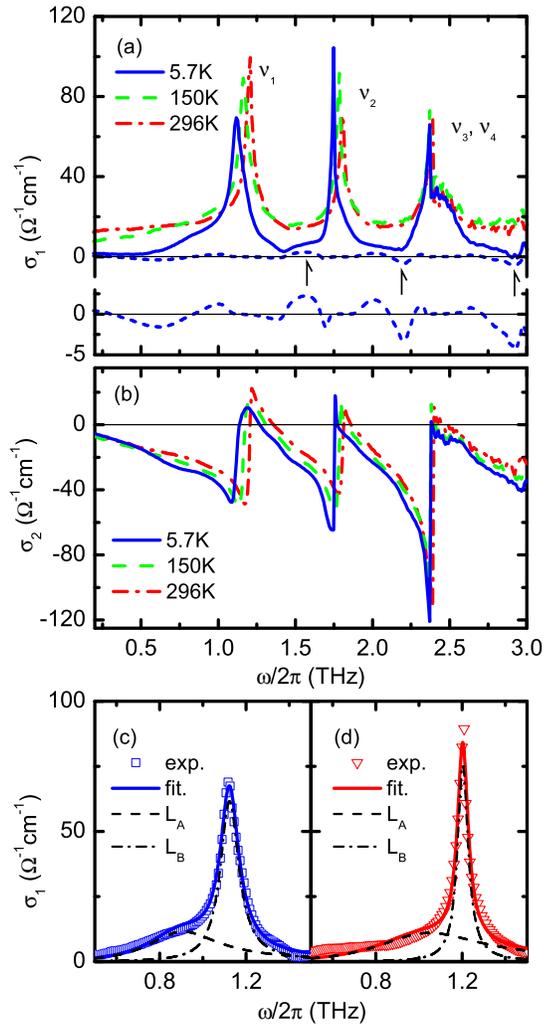}
\end{center}
\caption{(color online)
(a) Real parts $\sigma_1 (\omega)$ and (b) imaginary parts $\sigma_2 (\omega)$ of complex conductivity spectra of $p$-type BGG.
Dotted line and lower panel in (a) display the estimated interference pattern as described in Appendix B.
(c) and (d): Phonon spectra for rattling mode $\nu_{1}$ at 5.7\,K and 296\,K.
Experimental data are shown by open symbols and the fitting curves with two Lorentzians by solid lines, see the text.
For convenience, 1\,THz = 33.3\,cm$^{-1}$ = 48\,K = 4.14\,meV.}
\label{fig1}
\end{figure}

Figure \ref{fig1} (a) and (b) show the real part $\sigma_1(\omega)$ and the imaginary part $\sigma_2 (\omega)$ of the complex conductivity, respectively, 
at temperatures 296\,K, 150\,K and 5.7\,K. 
The $\sigma_1(\omega)$ spectrum presents distinct peaks superimposed on almost constant conductivity due to charge carriers, 
whereas $\sigma_2 (\omega)$ shows the corresponding frequency derivatives, as previously observed \cite{Mori2009}. 
The low-lying spectra observed around 1.2 and 1.8\,THz, assigned as $\nu_1$ and $\nu_{2}$, show a drastic change with decreasing temperature, associated with the phonon softening. 
In contrast, the higher frequency peaks $\nu_{3}$ (a collective cage mode) and $\nu_{4}$ (a Ba(1) mode) around 2.4\,THz are weakly dependent on temperature.

As indicated by arrows in Fig. \ref{fig1} (a), we have identified a few odd structures most clearly seen in the 5.7\,K spectrum; 
for example a hump around 1.5\,THz and dip around 2.2\,THz. 
Those small structures are attributed to an interference effect between multiply reflected lights in the present highly dielectric material (See also Ref. \cite{Mori2009}). 
The dotted line in Fig. \ref{fig1} (a) shows the interference pattern, which is estimated as described in detail in Appendix B. 
In the lower panel of Fig. \ref{fig1} (a), the interference pattern is also shown by an expanded scale. 
Those spurious effects are observed weakly but clearly for frequency regimes in between phonon peaks or at low temperatures where any carrier contributions are hardly traced. 
We should mention here that a broad sideband at the lower frequency side of the $\nu_1$ peak still appears even after the background correction is made. 
Since such a structure is reproducibly observed in our measurements for other samples with different thicknesses, we conclude that this remaining sideband structure is intrinsic. 
For detailed discussions, see also Appendix B. 

From the corrected spectra, the temperature-dependent, non-Drude-like contributions of charge carriers are 
subtracted by assuming a linear dispersion with some gradient as an adjustable parameter for the sum rule to hold for oscillator strengths. 
In Fig. \ref{fig1} (c) and (d) we show the peak profiles for the rattling $\nu_{1}$ mode, respectively at 5.7\,K and 296\,K.

\begin{figure}
\begin{center}
\includegraphics[width=6.5cm]{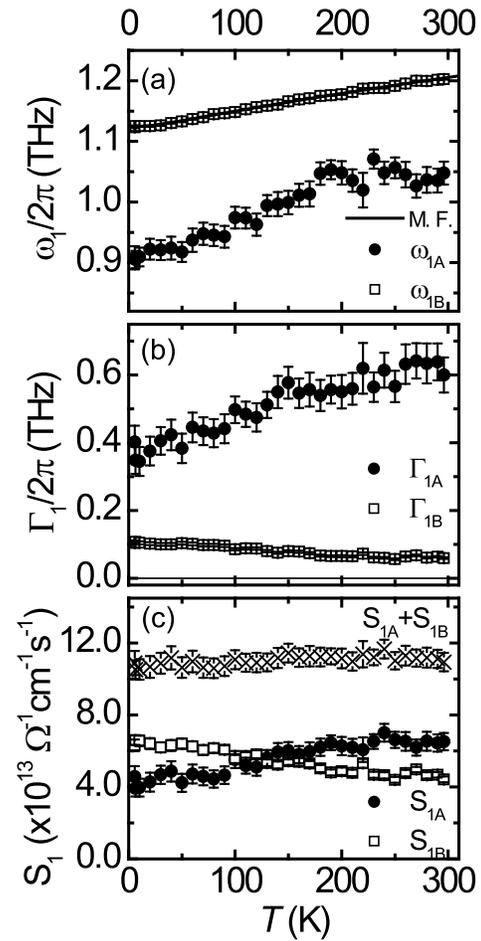}
\end{center}
\caption{(color online)
(a) Temperature dependence of the peak frequency $\omega_{1 \rm A}$ (filled circles) and $\omega_{1 \rm B}$ (open squares).
The theoretical curve obtained from the anharmonic potential model \cite{Dahm2007, Matsumoto2009} is also plotted as a solid line.
(b) Temperature dependence of the linewidth $\Gamma_{1 \rm A}$ (filled circles) and $\Gamma_{1 \rm B}$ (open squares).
(c) Temperature dependence of the oscillator strength $S_{1 \rm A}$ (filled circles) and $S_{1 \rm B}$ (open squares).
The total oscillator strength $S_{1 \rm A} + S_{1 \rm B}$ (crosses) are also plotted.
}
\label{fig2}
\end{figure}

To see these spectra in more detail, we fit the data by the sum of two Lorentzian curves $L_{\rm A}$ and $L_{\rm B}$, 
\begin{eqnarray}
L_i (\omega) = \frac{(2/\pi) S_{1i} \omega ^2 \Gamma _{1i}}{(\omega _{1i}^2 - \omega ^2)^2 + \omega ^2 \Gamma _{1i}^2 } ~(i = {\rm A, B}), 
\label{fitlor}
\end{eqnarray}
whereby the fitting parameters are the resonant frequencies, $\omega_{1 \rm A}$ and $\omega_{1 \rm B}$, the linewidths $\Gamma_{1 \rm A}$ and $\Gamma_{1 \rm B}$, 
and the oscillator strengths $S_{1 \rm A}$ and $S_{1 \rm B}$. 
As shown in Fig. \ref{fig2} (a), $\omega_{1 \rm A}$ and $\omega_{1 \rm B}$ decrease, respectively, by 10\,\% and 7\,\% from room temperature down to 5.7\,K, 
the latter of which closely reproduces our previous results \cite{Mori2009}. 
The temperature dependence is well explained by the model calculations for anharmonic vibrations \cite{Dahm2007,Matsumoto2009}. 
On the other hand, for the linewidth shown in Fig. \ref{fig2} (b), $\Gamma_{1 \rm A}$ decreases monotonically from 0.6\,THz at 296\,K to 0.4\,THz at 5.7\,K, 
whereas $\Gamma_{1 \rm B}$ of 0.06\,THz at 296\,K (consistent with INS measurements \cite{Christensen2008}) increases almost by a factor of 2, to 0.11\,THz at 5.7\,K.
Such an anomalous broadening can also be discussed from the difference between frequencies where $\sigma_2 (\omega)$ has a maximum and a minimum around 1.2\,THz, 
which is gradually enhanced toward low temperature, as seen in Fig \ref{fig1} (b). 
It is worth noting here that the ratio $S_{1 \rm B}/S_{1 \rm A}$ increases from 0.8 at 296\,K to 1.9 at 5.7\,K as seen in Fig. \ref{fig2} (c),
whereas the total contribution $S_{1 \rm A}+S_{1 \rm B}$ (marked by crosses) is conserved within this temperature range. 
Such a shift of optical weights indicates that $\nu_{1 \rm A}$ and $\nu_{1 \rm B}$ are not independent vibrational modes, since those do not satisfy the optical sum rule individually. 
From these readdressed data, we conclude that the main sharp peak of the $\nu_1$ mode is superimposed on a broad spectra weighted in the lower frequency regime 
and becomes broadened toward lower temperature.

\section{Discussions}
\subsection{Model for hybridized phonon system}
\begin{figure}[tb]
\begin{center}
\includegraphics[width=0.48\textwidth ]{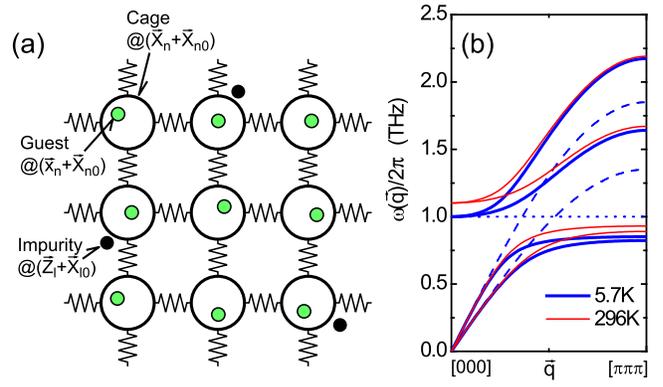}
\end{center}
\caption{(color online)
(a) Schematic representation of a rattling and acoustic phonon system in the cagelike materials.
(b) Zero-th order dispersion relation of the phonons propagating along the (111) direction, calculated from Eq. (\ref{Hamil0th}).
The rattling and acoustic branches without any hybridization are also plotted in dotted and broken lines, respectively .
}
\label{fig3}
\end{figure}

Here we propose an impurity-scattering model for the hybridized phonon system to interpret the observed anomalies in $\sigma_1 (\omega)$. 
As already introduced, the rattling and acoustic phonons are composed of an array of electronegative cages and electropositive guest ions. 
We consider a simplified model where the cages in the unit cell are represented by a single cage at each lattice position, as schematically illustrated in Fig. \ref{fig3} (a). 
We denote the lattice position by $\vec X_{n0}$, which is the equilibrium position of the $n$-th cage. 
The mass of the cage and the guest ion are $M_X$ and $m_x$, respectively. 
Then the field variables in the phonon system are given by the coordinate of the cage, $\vec X_{n0} + \vec X_n$ (the canonical momentum $\vec P_{Xn}$) 
and the coordinate of the guest ion, $\vec X_{n0} + \vec x_n$ (the canonical momentum $\vec p_{xn}$).

Applying an elastic theory to the cage system and a local potential model to the guest ion, we consider the following Hamiltonian,
\begin{align}
H_0 = & \sum _{n} \left[ \frac{\vec P_{Xn}^2}{M_X} + \frac{K_1}{2} \sum _{i = x,y,z} (\nabla _{i} {\vec X_{n}})^2 + \frac{K_2}{2} (\vec \nabla \cdot {\vec X_{n}})^{2}  \right] \notag \\
 & + \sum _{n} \left[ \frac{\vec p_{xn}^{\,2}}{m_x} + V(\vec r_n) \right],
\label{Hamil0th}
\end{align}
where $\nabla _{i}$ represents the gradient along the $i$ axis, and $\vec r_n = \vec x_n - \vec X_n$. 
By use of a center of mass coordinate, $\vec R_n = (M_X \vec X_n + m_x \vec x_n)/(M_X + m_x)$ and the relative coordinate $\vec r_n$, 
the Hamiltonian (\ref{Hamil0th}) represents both the acoustic phonon system with mass $M = M_X +m_x$ and spring constants $K_1$ and $K_2$,
plus the rattling phonon system with reduced mass $m = m_x M_X/(M_X + m_x)$ in the potential $V (\vec r)$. 
The interaction between $\vec R_n$ and $\vec r_n$ is mediated by 
\begin{equation}
	\begin{pmatrix} \vec X_n \cr \vec x_n \cr \end{pmatrix}
	=\begin{pmatrix} 1 & - \eta_X \cr 1 & \eta_x \cr \end{pmatrix}
		\begin{pmatrix} \vec R_n \cr \vec r_n \end{pmatrix} 
\end{equation}
where $\eta_X = m_x/M$ and $\eta_x = M_X/M$.
A strong mixing between the rattling and acoustic phonons is expected and shown as solid lines in Fig. \ref{fig3} (b), which is consistent with Ref. \cite{Christensen2008}.

The local potential $V (\vec r)$ is assumed as
\begin{equation}
V (\vec r) = \sum_{i = x,y,z} \left( \frac12 k r_{ni}^2 + \frac14 \lambda r_{ni}^4 \right)
\end{equation}
for simplicity, as discussed in the one-dimensional anharmonic potential (1D-AP) model \cite{Dahm2007, Matsumoto2009}. 
We neglect the directional difference in the real cage, by assuming the averaged effect in the unit cell and concentrating on the lowest-lying rattling oscillation. 
We also assume a quasi-on-center rattling as already mentioned. 
To note, the linewidth at 296\,K is much smaller than that predicted from 1D-AP \cite{Mori2009,Matsumoto2009}. 
The rattling excitation may have a more propagating nature through the interaction with the acoustic phonon, 
and the local potential may contribute rather as a mean field, the energy-spread of the level transitions being washed away. 
The mean field approximation in Ref. \cite{Dahm2007},
\begin{equation}
m \omega_{\rm MF} ^2 \approx  k + 3 \lambda \langle r_x^2 \rangle,
\label{meanf}
\end{equation}
and the local potential approximation in Ref. \cite{Matsumoto2009} give the same result of frequency softening in the present parameter range. 
The theoretical curve in Fig. \ref{fig2} (a) is the model result with parameters $(1/2\pi)\sqrt{k/m} = 1.115\,{\rm THz}$ and $\hbar \lambda/\sqrt{m k^3} = 1.17 \times 10^{-2}$. 
The higher order contributions from $\lambda$ are negligible, so the Hamiltonian (\ref{Hamil0th}) gives no information on the linewidth.

The observed temperature dependence of the spectral linewidth motivates the identification of an origin for the anomaly. 
It does not show any increase with temperature, which would be expected from certain dissipative mechanisms into a continuum of multi-bosonic particle states. 
As previously mentioned, the charge carriers can have little effect in phonon scattering. 
The remaining possibility could reside in elastic scatterings from impurities or defects. 
Let us consider local potentials of rattling and acoustic phonons from impurities situated at $\vec Z_{\ell} + \vec X_{\ell0}$;
\begin{equation}
H_I = \sum _{\ell} \biggl[ \frac{U_{c}}{2}(\vec X_\ell - \vec Z_\ell)^2 + \frac{U_{g}}{2}(\vec x_\ell - \vec Z_\ell)^2 \biggr] ,
\label{HamilInt}
\end{equation}
where $U_c$ and $U_g$ are the force constants of impurity potentials for the cages and the guest ions, respectively.

Now the phonon retarded Green function, $G(\omega , \vec q)$ is obtained from
\begin{align}
G(\omega,\vec q)^{-1}  = & \left(
\begin{matrix}
\omega^2 - \omega^2 (\vec q) - \Pi_{RR}(\omega) & , \cr
\eta_X \omega^2 (\vec q) - \Pi_{Rr} (\omega) & , \cr
\end{matrix}
\right. \notag\\
 & \left.
\begin{matrix}
\eta_X \omega^2 (\vec q) - \Pi_{rR} (\omega) \cr
\omega^2 - \omega_{MF}^2 - \eta_X^2 \omega^2 (\vec q) - \Pi_{rr} (\omega) \cr
\end{matrix}
\right ) .
\label{Gph}
\end{align}
For simplicity, we assume the separation of the longitudinal and transverse components, indicated by the suffixes $L$ and $T$ respectively, and a simple cubic lattice giving
\begin{equation}
\omega _{T,L}^2 (\vec q) = \frac12 \omega _{T,L}^2 \left[ 1 - \frac13 (\cos q_x a + \cos q_y a + \cos q_z a) \right],
\end{equation}
where $\omega_T = \sqrt{12 K_1/M}$, $\omega_L = \sqrt{12 (K_1 + K_2)/M}$ and $a$ is the lattice constant. 
The latter assumption about $\omega _{T,L} (\vec q)$ is exact for the type-I clathrate structure. 
The self-energies obtained from the interaction (\ref{HamilInt}) $\Pi (\omega)$'s are renormalized as $\Pi(0)=0$, consistent with translational invariance, 
and are expressed as
\begin{align}
\lefteqn{ 
\begin{pmatrix} \Pi_{RR} (\omega) & \Pi_{rR} (\omega) \cr
\Pi_{Rr} (\omega) & \Pi_{rr} (\omega) \cr \end{pmatrix} =
n_i \begin{pmatrix} 1 & 1 \\ -\eta_X & \eta_x \\ \end{pmatrix} \times} & \notag \\
 & \begin{pmatrix}
U_{c}^2 G_{XX}(\omega) & U_{c}U_{g} G_{Xx}(\omega) \\
U_{c}U_{g} G_{xX}(\omega) & U_{g}^2 G_{xx}(\omega) \\
\end{pmatrix}
\begin{pmatrix} 1 & -\eta_X \\ 1 & \eta_x \\ \end{pmatrix}. 
\label{selfE}
\end{align}
Here $n_i$ is the density of impurities, 
and the equipositional Green functions for $\vec X_n$, $G_{XX} (\omega)$, and $\vec x_n$, $G_{xx} (\omega)$, etc, are obtained as 
\begin{align}
\lefteqn{\begin{pmatrix}
G_{XX}(\omega) & G_{Xx}(\omega) \\
G_{xX}(\omega) & G_{xx}(\omega) \\
\end{pmatrix}}  \notag \\
& =\frac{a^3}{(2\pi)^3} \int_{\Omega_B} d^3q
\begin{pmatrix} 1 & -\eta_X \\ 1 & \eta_x \\ \end{pmatrix}
G(\omega, \vec q)
\begin{pmatrix} 1 & 1 \\ -\eta_X & \eta_x \\ \end{pmatrix}
\label{Gxx}
\end{align}
where $\Omega_B$ is in the first Brillouin zone. 
By iterating the equations for the phonon propagators (\ref{Gph}), the mean field (\ref{meanf}) and the self energy (\ref{selfE}), 
we can obtain the phonon Green functions and the optical conductivity self-consistently. 
Even when there are several cages in the unit cell, Eq. (\ref{Gph}) is not much modified, 
if restricted to the lowest rattling vibration and the acoustic phonon. Details of the formulation will be presented elsewhere.

\subsection{Calculated optical conductivity and discussions}
According to linear-response theory, the optical conductivity $\hat \sigma (\omega)$ is related to the Green function at the $\Gamma$ point for $\vec r_n$ as
\begin{equation}
\hat \sigma (\omega) = i\omega \frac{Nq^2}{\hbar} G_{rr} ( \omega, \vec q = 0),
\label{linres}
\end{equation}
because only the relative motion between the cage and the guest ion is responsible for the induced electric field. 
Here, $N$ is the density of guest modes, $q$ is the effective charge, and $G_{rr} (\omega, \vec q)$ is an $r$-$r$ component of $G(\omega,\vec q)$.

\begin{figure}[tb]
\begin{center}
\includegraphics[width=0.48\textwidth ]{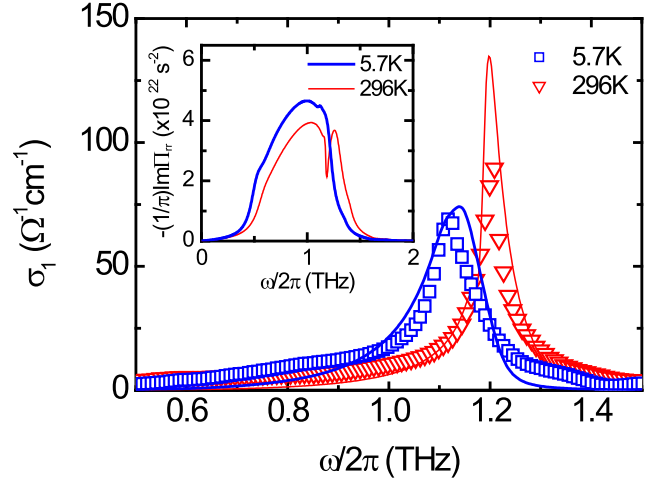}
\end{center}
\caption{
Spectral shapes of $\sigma _1(\omega)$ from $\nu_1$ at 5.7\,K and 296\,K.
The open symbols and solid lines indicate the data and calculations, respectively.
The calculated self-energy $-(1 / \pi ) {\rm Im} \Pi_{rr}$ at 5.7\,K and 296\,K are also shown in the inset.
Details are described in text.
}
\label{fig4}
\end{figure}

For the numerical calculation of $\sigma_1(\omega)$ for the $\nu_1$ mode, we fix the parameters as follows. 
By using the sound velocities from the INS measurements \cite{Christensen2008}, 
the frequency of the transverse and longitudinal acoustic modes at the zone boundary are estimated as 
$\omega_T/2\pi = 1.35\,{\rm THz}$ and $\omega_L/2\pi = 1.85\,{\rm THz}$, respectively. 
For the rattling phonons, we adopt $\omega_0/2\pi = 1.0\,{\rm THz}$, and $\bar \lambda = 1.4 \times 10^{-2}$, 
where $\omega_0 = \sqrt{k/m}$ and $\bar \lambda = \hbar \lambda/m^2 \omega_0^3$ are the resonant frequency and the dimensionless anharmonicity parameter, respectively. 
Taking into account the stoichiometry of type-I BGG, the mass ratio is assumed as $m_x/M_X =  1/3$. 
The coupling constants are $n_i U_{g}^2/K_1^2 = n_i U_{c}^2/K_1^2 = 1.73 \times 10^{-3}$. 
With these parameters the calculations are compared with the observed spectra as shown in Fig. \ref{fig4}. 
The experimental data, particularly the linewidth broadening toward low temperature and the lineshapes, can be well described by our calculations.

In order to understand the linewidth broadening at low temperature, the imaginary part ${\rm Im} \Pi _{rr}/\pi$ of the self-energy is shown in the inset of Fig. \ref{fig4}. 
As is obvious from Eq. (\ref{selfE}), the self-energy reflects the density-of-states (DOS) of rattling and acoustic phonons. 
Therefore, the rattling phonon spectra have a sideband structure reflecting the phonon DOS. 
Since the DOS is determined self-consistently, this process gives a strong self-interaction. 
One can see from the inset of Fig. \ref{fig4}, a large modification of the self-energy contribution specially around $\omega_{\rm MF}$. 
In the present model the temperature dependence comes only through $\omega_{\rm MF}$, which shows softening toward low temperature due to the anharmonicity effect. 
Strongly hybridized rattling and acoustic phonons are responsible for a large contribution to the DOS at the zone boundary. 
Since the frequency of the rattling phonon is low enough to hybridize with the acoustic phonon as the INS measurement have revealed \cite{Christensen2008}, 
a large DOS renormalization occurs around the frequency of the rattling phonon. 
When the temperature is lowered, $\omega_{\rm MF}$ is softened, which bends down the hybridized acoustic branch and increases the DOS at the zone boundary. 
This gives a large dissipation in the rattling phonon, filling the would-be hybridization gap and resulting in wider linewidth at low temperature. 
Both the softening due to anharmonicity and the strong hybridization between rattling and acoustic phonons are responsible for the linewidth broadening toward low temperature.

The above discussions indicate that not only the hybridization of the acoustic and rattling phonons 
but also impurity scattering in the hybridized phonon system is significant for the dynamical properties of clathrate compounds. 
As is obvious from Eq. (\ref{selfE}), the impurities can also dissipate the heat-conducting acoustic phonons around $\omega_{\rm MF}$ strongly through the large DOS. 
Here, the origins of such impurity scattering will require further investigation but one can consider several candidates, 
such as vacancies and/or substitutions of Ge by Ga in the framework. 
The contribution of impurities should also be taken into account for the heat transport behavior in this system.

Although we have not discussed any multiphonon scattering including the Umklapp process, 
their importance should not be neglected for the suppressed thermal conductivity in the hybridized phonon system. 
When the two phonons $\vec q$ and $\vec{q'}$ scatter and the momentum of the resulting phonons $\vec{q''}$ is outside the Brillouin zone, it is an Umklapp process. 
Since the Umklapp scattering can take place only when phonons have propagation wavevector $\vec{q'}$ and $\vec{q''}$ around the zone boundary,
such a process may not significantly affect the rattler modes at $\vec q \simeq 0$. 
Further investigations on the phonon dispersion relation and its temperature dependence, based on INS and/or Brillouin scattering measurements, will be required.

\section{Conclusion}
We have performed a detailed experiment of the optical conductivity measurements for type-I BGG having $p$-type carriers by THz-TDS with improved frequency resolution 
and have further clarified the dynamical properties of the infrared-active rattling phonons. 
Ba(2)$^{2+}$ modes in the oversized cage show softening and anomalous broadening toward low temperature. 
They also have broad sideband structures at the lower frequency part of the sharp main peak. 
Our discussion, based on an impurity scattering model for rattling modes that are strongly hybridized with cage acoustic phonons, 
offers and interpretation whereby the anomalous features of the rattling phonon spectra arise from both the softening 
due to anharmonicity and the strong hybridization between rattling and acoustic phonons.

We also point out that the present experiment has disclosed that the anharmonic potential for the rattling phonons works as a mean field at room temperature 
due to the strong hybridization with the acoustic phonon, and the perturbational treatment is justified. 
This result raises an interesting problem in the off-center potential, for example in BGS, 
since a crossover from a mean field regime to a degenerate level transition is expected with decreasing temperature. 
In this sense, interplays between carrier density and potential type remains as future problems in order to clarify the mechanism operating in cage systems. 

\begin{acknowledgment}
\section*{Acknowledgment}
The present works have been supported by the Global COE program ``Materials Integrations'', Grants-in-Aid for Scientific Research (A)(15201019), (A)(18204032) and (C)(23500056), the priority area ``Nanospace''(1951011) and ``Skutterudite''(15072205), the innovative areas ``Heavy Electrons''(20102004) from MEXT, Japan, and the Sasakawa Scientific Research from Japan Science Society.
\end{acknowledgment}

\appendix
\section{Frequency resolution}

\begin{figure}[t]
\begin{center}
\includegraphics[width=8cm]{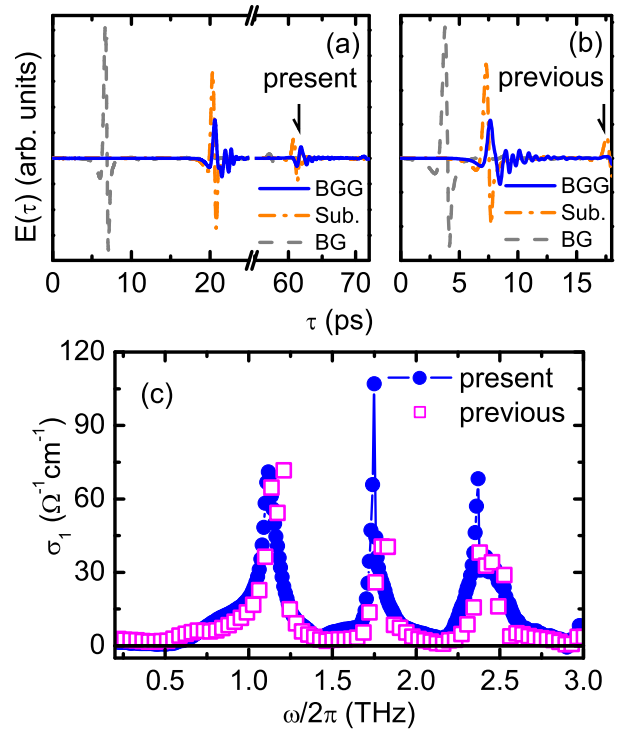}
\end{center}
\caption{(color online)
(a, b) Time evolution of THz-wave electric fields at low temperature obtained from three different sets of measurements (a) in present work and (b) in previous work \cite{Mori2009};
vacuum (broken lines, BG), the substrate only (dot-dashed lines, Sub.), and the type-I Ba$_8$Ga$_{16}$Ge$_{30}$ (solid lines, BGG) sample glued onto the substrate (solid lines). 
The second pulses reciprocated in a substrate are indicated by arrows. 
(c) The real part, $\sigma_1 (\omega)$, of complex conductivity at low temperature in the present work (filled symbols) compared with that in the previous work\cite{Mori2009} 
(open symbols).}
\label{fig0}
\end{figure}

The time evolution of electric fields $E(\tau)$ at low temperature measured in the present and previous works\cite{Mori2009},
obtained from three different sets of measurements, vacuum (BG), the substrate only (Sub.), 
and the sample glued onto the substrate (BGG), are shown in Fig. \ref{fig0} (a) and (b), respectively. 
As indicated by arrows, the second pulses reciprocated in a substrate appear only in $E(\tau)$ of the substrate and sample, 
which delay, respectively, about 40\,ps and 10\,ps behind the first pulse. 
To obtain the Fourier-transformed spectra, $E(\tau)$ after the second pulses must be {\it zero-filled} by multiplying the following rectangular window function
\begin{equation}
\Theta _T (\tau) = \left \{ 
\begin{array}{ll}
1 & (0 \leq \tau \leq T) \\
0 & ({\it otherwise}), \\
\end{array}
\right. 
\end{equation} 
where $T$ is the effective length of time delay between the first and second pulses. 
Since the resolution functions, which are Fourier components of $\Theta _T (\tau)$, are convoluted into the Fourier-transformed spectra, 
the actual frequency resolution is limited by the thickness of a substrate. 
Thus, by using a thicker substrate, the frequency resolution can be improved. 
Figure \ref{fig0} (c) shows the real part $\sigma_1 (\omega)$ of the optical conductivity obtained in the present work compared with that previously reported \cite{Mori2009}. 
The linewidth of each observed peak in the present work is narrower than that in previous one. 

\section{Analysis for interference effect}

The multiply reflected lights in a sample can interfere with each other, which often give rise to the spurious periodic pattern in the optical conductivity spectra. 
Thus, we obtain the optical conductivity spectra without an interference effect \cite{Dressel} 
between multiply reflected lights in the present highly dielectric material (See also Ref.\cite{Mori2009, Mori2011}) by the following procedures. 
First, the complex refractive index is numerically determined by using the equation for the transmission coefficient $t (\omega)$ 
taking into account multiple reflections in the sample and adhesive; 
\begin{equation}
\begin{split}
t (\omega) = & \frac{t_{\it vf} t_{\it fm} t_{\it ms}} {t_{\it vs}} \exp \{ i [\phi_{\it f} + \phi_{\it m} - \omega (d_f + d_g)/c_0] \} \\
& \times \{1 - r_{\it fm} r_{\it fv} \exp (2i\phi_{\it f}) - r_{\it ms} r_{\it gf} \exp (2i \phi_{\it m}) \\
& - r_{\it ms} r_{\it fv} \exp (2i \phi_{\it f} + 2i \phi_{\it m} ) \}^{-1}, \\
\end{split}
\label{eqnt}
\end{equation}
where $\hat n_i$ is the complex refractive index, $c_0$ the velocity of light, $\phi_i = \hat{n}_i \omega d_i/c_0$ the phaseshift, 
$t_{ij} = \frac{2\hat{n}_j}{\hat{n}_i + \hat{n}_j}$ and $r_{ij} = \frac{\hat{n}_i - \hat{n}_j}{\hat{n}_i + \hat{n}_j}$ are the complex transmission and reflection coefficients, respectively, 
and thickness $d_i$ of each medium are taken as $d_f \simeq d_m \ll d_s$. 
The subscripts {\it v, s, m, {\rm and} f} represent vacuum, substrate, adhesive, and sample, respectively. 
Here to note, the second term in the denominator of Eq. \ref{eqnt} describes the interference in the sample, 
the third one in the adhesive, and the forth one among the sample and adhesive, respectively. 
For numerical analyses of $\hat n_f$, we assumed $\hat n_m$ as a constant value of $\hat n_m = 1.66$, which was obtained from another THz measurement. 
Then, the complex conductivity $\hat \sigma$ is obtained from 
\begin{equation}
\hat n_f^2 = 1 + \frac{4 \pi i \hat \sigma }{ \omega }. 
\label{eqns}
\end{equation}

Next, we estimate the remaining interference effect due to the very slight misalignment from the ideal configuration as follows. 
This effect, which was observed sample-dependently as well in Fig. \ref{fig0} (c), as the small dip at 0.8\,THz for example, 
remain weakly in $\hat \sigma$ even though the multiple reflections in the sample are taken into account by using Eq. \ref{eqnt}. 
As one can see from Eq. \ref{eqnt}, an interference effect appears when the extinction coefficient $\kappa _f$ caused by phonons or electrons is small.
From the denominator of Eq. \ref{eqnt}, the remaining interference pattern $\Delta \kappa _f (\omega)$ in $\kappa _f$ of the sample is approximately proportional to  
\begin{equation}
\Delta \kappa _f (\omega) \propto \ln[1 - R \exp (-2 \kappa_f \omega d_f/c_0) \cos (2 n_f \omega d_f/c_0)], 
\end{equation}
when the refractive index $n_f$ of the sample is much larger than that $n_m$ of the adhesive and $\kappa _f \ll 1$ (where $R = |\hat n_f - 1|^2/|\hat n_f + 1|^2$ is the reflectivity). 
Then, the remaining interference pattern $\Delta \sigma _1 (\omega)$ in $\sigma _1$ is approximated to $ n_f \Delta \kappa \omega/ 2\pi$ from Eq \ref{eqns}, 
when $\Delta \kappa \ll 1$ and $\kappa _f \ll 1$. 
Finally, $\Delta \sigma _1 (\omega)$ is estimated from
\begin{equation}
\begin{split}
\Delta \sigma _1 (\omega) \sim & A_{\rm o}(n_f\omega/2\pi) \\
 & \times \ln[1 - R \exp (-2 \kappa_f \omega d/c) \cos (2 n \omega d/c)]. \\
\end{split}
\end{equation}
Taking into account the temperature- and frequency-dependent $\hat n_f$, 
these spurious effects are most consistently subtracted from the data by adjusting the amplitude parameter $A_{\rm o}$ of about 3.

\end{document}